
\input psfig

\overfullrule0pt
\def\MS{{\overline{\rm MS}}}
\def\O{{\cal O}}
\def\KS{KS}
\def\Pade{Pad\'e}

 \PHYSREV
\def\Buildrel#1\under#2{\mathrel{\mathop{#2}\limits_{#1}}}

\REF\refI{
M. A. Samuel, G. Li and E. Steinfelds, Phys. Rev. {\bf D48}, 869 (1993).}
\REF\refII{M. A. Samuel and G. Li,
Int. J. Th. Phys. {\bf 33}, 1461 (1994).}
\REF\refIII{
M. A. Samuel, G. Li and E. Steinfelds, Phys. Lett. {\bf B323}, 188
(1994).}
\REF\refIV{M. A. Samuel and G. Li,  Phys. Lett. {\bf B331}, 114 (1994).}
\REF\refV{M. A. Samuel, G. Li and E. Steinfelds,
``On Estimating Perturbative
Coefficients in QFT and Statistical Physics,''
Phys. Rev. E (to be published, 1995).}
\REF\refVI{
M. A. Samuel, ``On Estimating Perturbative Coefficients in QFT,
Statistical Physics and Mathematics,'' Oklahoma State University
Research Note 290, July (1994).}
\REF\refVII{
J. Ellis, M. Karliner, M. Samuel and E. Steinfelds, ``The
Anomalous Magnetic Moments of the Electron and the Muon-Higher
Precision with Pad\'e Approximants,'' SLAC-PUB-6670 (1994),
hep-ph/9409376.}
\REF\refVIII{
J. Ellis, M. Karliner and M. Samuel, ``Towards an Improved
Determination of $\alpha_s$ Using Higher-Order Perturbative QCD.''
(in preparation).}
\REF\refIX{
A. I. Vainshtein and V. I. Zakharov, Phys. Rev. Lett. {\bf 73}, 1207
(1994).}
\REF\refIXa{
C.N. Lovett-Turner and C.J. Maxwell,
Nucl. Phys. {\bf B432}, 147 (1994).
}
\REF\refX{
E. Braaten, Phys. Rev. Lett. {\bf 71}, 1316 (1993) and references
therein.}
\REF\refXI{
M. Davier, {\it Proc. 2-nd Workshop on $\tau$ Lepton
Physics,} Ohio State U., Columbus, Ohio, Sep.
8--11, 1992, pg. 514, K. K. Gan, Ed., World Scientific (1993).}
\REF\refXII{
W. Trischuk, {\it Proc. 2-nd Workshop on $\tau$ Lepton
Physics,}, {\it op. cit.},
pg. 59;
W. J. Marciano, {\it Proc. DPF-92
Meeting,} Fermilab, Nov. (1992).}
\REF\refXIII{
L. R. Surguladze and M. A. Samuel, in {\it Beyond the Standard Model
II,} Norman, Oklahoma, Nov. 1--3, 1990; K. Milton,
R. Kantowski and M. A. Samuel, Eds., World Scientific (1991);
L. R. Surguladze and M. A. Samuel,
Phys. Rev. Lett. {\bf 66}, 560 (1991); S. G. Gorishny, A. L. Kataev
and S. A. Larin, Phys. Lett. {\bf B259}, 144 (1991).}
\REF\refXIV{
J. J. Hernandez \etal, Phys. Lett. {\bf B239}, 1 (1990).
W. De Boer, ``Experimental Results on QCD from $\epem$ Annihilation,''
SLAC--PUB--4428 (1987), invited talk  at the {\it 10th
Warsaw Symposium on Elementary Particle Physics,} Kazimierz, Poland,
May 25--29, 1987, p 503.}
\REF\refXV{
G. D'Agostini, W. De Boer and G. Grindhammer, Phys. Lett. {\bf B229},
160 (1989).}

\REF\refXVI{
LEP Electroweak Working Group, CERN-PPE-94-187, Nov 1994,
Contributed to ICHEP 94, Glasgow, 1994.}

\REF\refXVII{
A. L. Kataev and V. V. Starshenko,
Mod. Phys. Lett. {\bf A10}, 235 (1995).}

\REF\refXVIII{
F. Le Diberder,
{\sl Experimental estimates of higher order perturbative corrections to
$R_\tau$},
Orsay preprint, LAL-94-43,
submitted to Phys. Lett. B.}

 \pubnum{6589}
 \date{July 1994}
 \pubtype{T/E}
\nopubblock
 \titlepage
\baselineskip=12pt
\line{\hfill}
\vskip-3cm
\line{\hfill SLAC-PUB-6589}
\line{\hfill TAUP-2240/95}
\line{\hfill CERN-TH/95-68}
\line{\hfill hep-ph/9503411}
\vskip 1cm
 \title{Comparison of the Pad\'e Approximation Method to Perturbative QCD
Calculations
 \doeack
\foot{To appear in Phys. Rev. Lett.}}
 \author{Mark A. Samuel}
 \address{Department of Physics\break
 Oklahoma State University\break
 Stillwater, OK 74078\break
e-mail: physmas@mvs.ucc.okstate.edu}
\andaddress \SLAC
\andauthor{John Ellis}
\address{CERN\break
Geneva\break
Switzerland\break
e-mail: johne@cernvm.cern.ch}
\andauthor{Marek Karliner}
\address{School of Physics and Astronomy\break
Tel--Aviv University\break
Tel--Aviv, Israel\break
e-mail: marek@vm.tau.ac.il}

 \medskip
\abstract
We present a method of estimating perturbative coefficients in
Quantum Field Theory using \Pade\ Approximants. We test this method on
various known QCD results, and find that the method works very well.
\hfill
\endpage

\doublespace
By using the first $n$ coefficients in a series expansion, we have
estimated the $(n+1)$-st perturbative coefficient in Quantum Field
Theory (QFT). Though there is currently no theoretical basis for
extrapolating coefficients in the perturbative loop expansion of
QFT by our method, our results have thus far been in good agreement
with the calculated coefficients of quantum electrodynamics (QED),
as well as with series in statistical physics, condensed matter
theory and mathematics [\refI-\refVI].
In this paper we compare our method to
the perturbative loop expansion of quantum chromodynamics (QCD)
at the five-loop level.
We shall present results for the $R$-ratio, the $R_\tau$ ratio,
the QCD $\beta$ Function and two QCD Sum Rules.

Our method makes use of Pad\'e Approximants (PA) and enables
us to obtain an estimate and an error-bar for each coefficient.
We define the PA
$$ [N/M] = {a_0+a_1X+\cdots+a_NX^N\over
1+b_1X+\cdots+b_MX^M} \eqn\I $$
to the series $S$ where we set
$$ [N/M] = S+ {\cal O}(X^{N+M+1})
\qquad\hbox{and}\qquad
S = \sum^\infty_{n=0} S_nX^n \ .
\eqn\II$$
One solves Eq. \II\ and then predicts the coefficient of the next
term $S_{N+M+1}$.  This is what we do in this paper.  One can also
use the full PA to estimate the sum of the whole series $S$. This is
what we will do in the future [\refVII,\refVIII].
For a detailed description see
Refs. [\refV] and [\refVI].

The PA's are known to accelerate the convergence of many series by
including the effects of higher (unknown) terms, thus providing a
more accurate estimate of the series.  We have recently proved the
following theorem, which provides a useful sufficient condition for
the PA's to be accurate.  Defining
$ f(n) \equiv \ell n\, S_n$
and considering
$$ g(n) = {d^2f(n) \over dn^2}
\eqn\V $$
a sufficient condition for the PA's to converge is that
$$\Buildrel  {n \rarrow \infty}\under\lim \ g(n) = 0 \ .
\eqn\VI $$
The PA's thus provide reliable estimates of asymptotic series whose
coefficients diverge as
$$ S_n = n!\, k^n n^\gamma \ ,
\eqn\VII $$
as is believed to be the case in QED and QCD [\refIX].  It can easily be
shown that Eq. \VI\ is satisfied for $S_n$ given by Eq. \VII.
In the cases of these and other series whose Borel transform
has a finite radius of convergence,
the higher-order PA's give progressively better approximations
to the Principal Value of the transform integral over
Borel singularities.
It is easy to check that for series with one or
two simple Borel poles, (i.e. IR and UV renormalons),
\Pade\ approximants predict the next term in a given series
with a rapidly increasing precision. In these
cases an analytic estimate $\sim M!/N^M$
can be made for the relative error
of the $[N/M]$ Pad\'e approximant prediction of the next term
in the series.

It might be objected, however,
that these cases are not sufficiently complicated to
be realistic. Therefore, as an exercise, we have
evaluated Pad\'e approximants to the large-$N_f$ limit of
the vacuum polarization D-function in QCD,
which is known to all orders in $\alpha_s$, and
whose Borel
transform contains an infinite series of double poles
at both positive and negative integers.\refmark{\refIXa}
Once again, as seen in the Figure, the
Pad\'e approximants' predictions of the next term in the series
converge rapidly,
in agreement with the above-mentioned estimated error for
the $[N/M]$ Pad\'e approximant.
The convergence of the Borel transform of the D-function
series, in particular, indicates that our PA approach is
well suited for perturbation series with the asymptotic
behavior expected in QCD.

\vbox{
\vskip0.8cm
\centerline{\psfig{figure=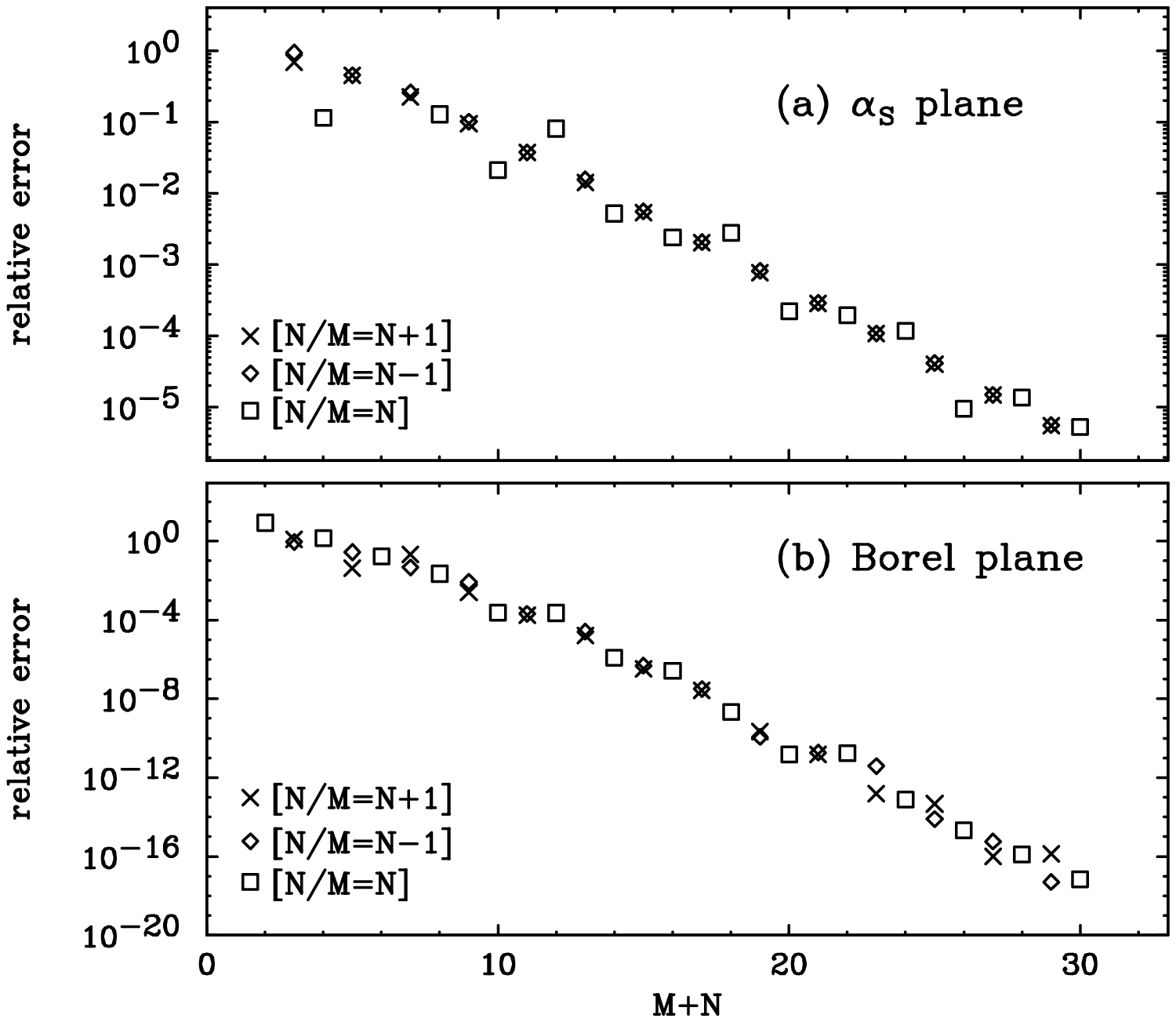,height=5in}}
\vskip0.5cm
\vbox{\tenpoint\baselineskip=14pt
\noindent
Relative errors in the $[N/M]$ Pad\'e approximants:
(a) to the
QCD vacuum polarization D-function, evaluated to all orders
in the large-$N_f$ approximation\refmark{\refIXa} --
the rate of convergence agrees with expectations for
a series with a discrete set of Borel poles,
and
(b) to the Borel transform of the D-function series,
where the convergence is particularly striking.
}}

We now turn to some QCD applications of our PA approach.
Let us first consider the $R_\tau$ ratio\refmark{\refX}\
where $R_\tau$ is defined as follows:
$$ R_\tau \equiv \Gamma(\tau \rarrow \nu + \hbox{hadrons})/
\Gamma(\tau \rarrow e \nu \bar \nu) = 3S_{\rm EW}(r_\tau+r^1_\tau)
\eqn\VIII$$
where
$ S_{\rm EW} = 1.019 $
is the electroweak correction and
$ r^1_\tau = - 1.58\% $
is the non-perturbative contribution.
The perturbative QCD contribution is given by
$$ r_\tau = 1 + {\alpha_s\over\pi}+ 5.202\,
\left(\alpha_s\over\pi\right)^2+ 26.36\,
\left(\alpha_s\over\pi\right)^3+ (109.2\pm 12.9)\,
\left(\alpha_s\over\pi\right)^4
\eqn\XI$$
where the last coefficient is our estimate of the five-loop
contribution.
Using the experimental average\refmark{\refXI,\refXII}\
value
$   R_\tau = 3.623(17) $,
we obtain for the strong coupling constant
$$  \alpha_s(M_\tau) = 0.325(6) \ .
\eqn\XIII$$
where the error does not include various systematic
uncertainties, which
go beyond the scope of this letter and
 are discussed elsewhere.\refmark{\refVIII}
Numerically Eq. \XI\ becomes
$   r_\tau = 1+0.1035+0.0557+0.0292+0.0125{=}1.201(19)\ .$
One can see that the perturbative series is converging,
albeit somewhat slowly.
Using the known $\beta$ function (see Table II for our estimate of
the four-loop $\beta$ function), Eq. \XIII\ corresponds
to
$ \Lambda^{(3)} = 355(11)$  MeV
in the $\MS$ Scheme.
Stepping up through the 4 and 5 fermion-thresholds
($m_c=1.5$ GeV and $m_b=5$ GeV, respectively) we get
$ \Lambda^{(4)} = 306(11)$ MeV
and
$ \Lambda^{(5)} = 218(9)$ MeV,
and, hence,
$$ \alpha_s(34\ \hbox{GeV}) = 0.1399(11) \ .
\eqn\XVIII$$

For the $R$ ratio, we have\refmark{\refXIII}\
$$  R = 3\Sigma Q^2_f r\quad:\quad
 r = 1+\left(\alpha_s\over\pi\right)+1.409\,
\left(\alpha_s\over\pi\right)^2-12.805\,
\left(\alpha_s\over\pi\right)^3- (87.5\pm10.8)\,
\left(\alpha_s\over\pi\right)^4 \ .
\eqn\XX$$
The last term in Eq. \XX\ is our estimate for the five-loop
contribution to $R$ where we here extrapolated the related Adler
D-function.
Using Eq. \XVIII\ we obtain
$$ r(34\ \hbox{GeV}) = 1+0.0445+0.0028-0.0011-0.0003= 1.0459(4) \ .
\eqn\XXI$$
This series for $r$ in Eq. \XXI, where the contributions
in each order, up to five loops, are given, seems to converge nicely.
Experimentally there are two measurements of $r(34\  \hbox{GeV})$.
They are
$ r(34\ \hbox{GeV}) = 1.049(7)$\refmark{\refXIV}
and
$ r(34\ \hbox{GeV}) = 1.056(8)$\refmark{\refXV}.
It can be seen that the extrapolation prediction in
Eq. \XXI\ is in good agreement with these experimental values.
We can now evolve $\alpha_s$ up to $M_Z$, the $Z$ boson mass.
Our result is
$$  \alpha_s(M_Z) = 0.119(2) \ .
\eqn\XXIV$$
which is consistent with the latest experimental value obtained from
total cross-section measurements at LEP\refmark{\refXVI}
$$ \alpha_s(M_Z) = 0.126(6) \ .
\eqn\XXV$$
For $r$ at $M_Z$ we get
$ r(M_Z) = 1+0.0378+0.0020-0.0007-0.0002 = 1.0389(2) \ .$

We now present our estimates for
higher-order perturbative coefficients for
$R$, $R_\tau$, the QCD $\beta$
Function and two QCD Sum Rules. All of the results presented here are
in the $\MS$ scheme. The corresponding results for other schemes can
also be obtained. However the scheme-dependence should decrease in
higher order, disappearing if all orders are known.

To provide systematic-error estimates, we first apply our method to
$S_n$, obtaining our estimate $S^{(0)}_{n+1}$.  We then apply it to
the reciprocals $r_n \equiv 1/S_n$, and take the reciprocal
again to obtain
$S^{(1)}_{n+1}$.  We then take differences
$ t_n = r_{n+1}-r_n$
and apply our method to obtain a third estimate $S^{(2)}_{n+1}$.
Our systematic-error estimate is $\Delta/2$, where
$ \Delta = \left|S^{(2)}_{n+1}-S^{(1)}_{n+1}\right| \ .  $
We then combine the diagonal and non-diagonal estimates
of $S_{n+1}^{(0)}$, weighted by $1/\Delta^2$,
to obtain our final estimate for $S_{n+1}$.

In Table I we present our results for $R$ and $R_\tau$.
The first entry in each case is the four-loop result.
The light-by-light contribution is small but should be added
to $R(\MS)$.
Our estimates
of the four-loop coefficients, based on PA's to lower-order
coefficients,
agree well with the known exact results,
providing a sound phenomenological footing for our method.
The second entries are our estimates for the five-loop
coefficients and
the numbers in brackets are our estimate of the systematic errors.
The results $-$96.8 \KS\ and 105.5 \KS\ are the estimates
of Ref.~[\refXVII],
obtained using a completely independent method,
Optimized Perturbation Theory (OPT).
The agreement with our estimates is very good.

In Table II we present our results for the QCD $\beta$-function.
The agreement with the known 3-loop results is very good, and we
present the first estimates of the four-loop QCD $\beta$ function.
Note that the three- and four-loop results are scheme-dependent,
and we use the $\overline{MS}$ scheme throughout.  The 3-loop
result is the same for any MS-type scheme.  Since we use the
$\overline{MS}$ result for the 3-loop coefficient our estimate for the
four-loop coefficient is also for the $\overline{MS}$ scheme.

In Table III we present our results for the Bjorken sum rule for
deep inelastic (unpolarized) neutrino-nucleon scattering.
To compare with Ref.~[\refXVII],
we multiply their results BjnSR by $-$2/3.
Our results for $\O(\alpha^3_s)$ are very good again, and
again the agreement with Ref.~[\refXVII] in $\O(\alpha^4_s)$ is
excellent.
We present our results for the Bjorken sum rule for deep inelastic
polarized electron-nucleon scattering, $-$BjpSR in Table IV.
Again our results in $\O(\alpha^3_s)$ are good and our
$\O(\alpha^4_s)$ estimates agree with Ref.~[\refXVII].
The Gross-Llewellyn Smith sum rule (GLSSR) differs from the BjpSR by
the light-by-light contribution:
$    GLSSR = BjpSR - 0.413\, f \ .$
The light-by-light contribution here and for $R$ should be treated
separately.
This contribution is small, however, for all cases of interest.

In Table V we present our results for the $R$ ratio for various
$N_f$.
The small difference for $N_f=5$ compared with Table I is due to a
slight difference in averaging the various estimates.
The results for the four-loop coefficients are excellent and the
agreement with Ref.~[\refXVII]
for the five-loop coefficients is satisfying.

Although it is interesting that our estimates for the next term
agree with Ref.~[\refXVII] and we believe both of us are right,
we cannot be
certain.  Both of us may be wrong.  Fortunately, even a crude
estimate of the next term is sufficient since their contributions to
the full series are small.  However, it is important to know they are
small!

In conclusion, we have used our estimation method, which makes use
of Pad\'e Approximants, to estimate various perturbative
coefficients in QCD. Our estimates for the known terms is very good.
Moreover our estimates for the next unknown terms agree very well
with the results of Ref.~[\refXVII]
in all cases where comparison is possible.

\noindent\undertext{Note added:}\nextline
A phenomenological extraction of the five-loop D-function
coefficient
from measured  moments of $\tau$ decay data has recently
appeared\refmark{\refXVIII}, which confirms our prediction.

In the process of our analysis we are comparing our results with
explicit expressions given in ref. 14 and in papers listed in ref. 18.
The original references for the NNLO calculations are: $R(s)$ - Ref.
14 and S.G.Gorishny, A.L.Kataev, S.A.Larin in "Standard Model and
Beyond: from LEP to UNK and LHC" Dubna, USSR, October 1-5,1990,
S. Dubnicka, et al., Eds.  World Scientific (1991) p.288; the QCD
beta-function - O.V. Tarasov, A.A. Vladimirov and A.Yu. Zharkov,
Phys. Lett. B93(1980)429; the Bjorken non-polarized sum rule - S.A.
Larin, F.V. Tkachov and J. Vermaseren, Phys. Rev. Lett. 66(1991)862
and Bjorken polarized sum rule - S.A.Larin and J. Vermaseren, Phys.
Lett. B259(1991)345.  We thank Andrei Kataev for correspondence on
this matter.

\ack

We would like to thank the theory group at SLAC for its kind
hospitality. We would also like to thank the following people for
very helpful discussions:  David Atwood, Richard Blankenbecler, Stan
Brodsky, N. Deshpande, JoAnne Hewett, Helen Perk, Jacques Perk,
Dominique Pouliot, Helen Quinn, Tom Rizzo, Davison Soper, Levan
Surguladze and N.V.V.J. Swamy. This work was supported by the U.S.
Department of Energy under grant numbers DE-FG05-84ER40215 and
DE-AC03-76SF00515.
The research of M.K.
was supported in part
by grant No.~90-00342 from the United States-Israel
Binational Science Foundation (BSF), Jerusalem, Israel,
and by a Grant from the G.I.F., the
German-Israeli Foundation for Scientific Research and
Development.
\bigskip

\refout
\endpage

\overfullrule0pt
\baselineskip 14pt
\line{\hfill}
\vskip-1cm
\vbox{
\vbox{
\centerline{TABLE I}

\noindent
Estimates for $R(\MS),\ N_f=5$, and
$R_\tau(\MS),\ N_f=3$, at the four- (first row) and
five-loop (second row)
order.
The numbers in brackets are the estimated 1$\sigma$ error-bars.
The four-loop results are compared with the exact (known) results and
the five-loop results are compared with those of \KS.
$N_f$ is the number of fermions (quarks).
\medskip
\centerline{\vbox{\halign{
# &\qquad \hfill # &\qquad \hfill # \cr
       SERIES   &                ESTIMATE    &             EXACT    \cr
\noalign{\medskip\hrule\medskip}
                &                            &                      \cr
$R(\MS),\ N_f=5$&             $-$10.20(1.53) &            $-$12.76  \cr
                &                            &                      \cr
                &             $-$87.5(10.8)  &   $-$96.8$\,\,$(\KS) \cr
                &                            &                      \cr
$R_\tau(\MS),\ N_f=3$
   &                27.06(6.77) &                26.37 \cr
                &                            &                      \cr
                &               109.2(12.9)  & 105.5$\,\,$(\KS)       \cr
\noalign{\medskip\hrule\medskip}
}}}}
\vskip1cm
\vbox{
\centerline{TABLE II}
\noindent
Estimates for the QCD $\beta$-%
function in three- (first row) and four-loop (second row) order.
The 3-loop results are compared with the exact (known) results.
$N_f$ is the number of fermions (quarks).
\bigskip
\centerline{\vbox{\halign{
# &\qquad \hfill # &\qquad \hfill # \cr
       SERIES     &              ESTIMATE  &                EXACT \cr
\noalign{\medskip\hrule\medskip}
                  &                        &                      \cr
QCD $\beta$ FCN   &                        &                      \cr
                  &                        &                      \cr
$N_f =3$          &          $-$455(228)  &               $-$644   \cr
                  &            $-$5920(1956) &                 ---  \cr
                  &                        &                      \cr
                  &                        &                      \cr
$N_f=4$           &            $-$316(158)   &                $-$406  \cr
                  &            $-$3058(875)  &                ---   \cr
                  &                        &                      \cr
                  &                        &                      \cr
$N_f=5$           &            $-$195(49)    &             $-$181  \cr
                  &            $-$845(105)   &                 ---  \cr
                  &                        &                      \cr
\noalign{\medskip\hrule\medskip}
}}}}
\vfill
}
\eject
\vfill
\bigskip

\vbox{
\line{\hfill}
\vskip-1cm
\vbox{
\centerline{TABLE III}
\noindent
Estimates for the Bjorken sum rule for deep inelastic (unpolarized)
neutrino-nucleon scattering.
The $\O(\alpha^3_s)$  (first row) results are compared with the exact (known)
results and the $\O(\alpha^4_s)$ (second row) results  are
compared with those of
\KS.
We multiply their results by $-$2/3 for ease of comparison.
\bigskip
\centerline{\vbox{\halign{
# &\qquad \hfill # &\qquad \hfill # \cr
       SERIES    &               ESTIMATE &                 EXACT  \cr
\noalign{\medskip\hrule\medskip}
                 &                        &                        \cr
 BjnSR $X-2/3$  &                        &                        \cr
                 &                        &                        \cr
 $N_f=3$         &          $-$13.0(6.0)  &                $-$18.6 \cr
                 &         $-$116(69)     &         $-$133$\,\,$(\KS) \cr
                 &                        &                        \cr
                 &                        &                        \cr
 $N_f=4$         &        $-$10.5(4.7)    &               $-$13.4  \cr
                 &         $-$67.2(17.7)  &        $-$75.8$\,\,$(\KS) \cr
                 &                        &                        \cr
                 &                        &                        \cr
 $N_f=5$         &         $-$8.3(3.6)    &                -8.5    \cr
                 &         $-$30.6(1.4)   &        $-$29.4$\,\,$(\KS) \cr
                 &                        &                        \cr
\noalign{\medskip\hrule\medskip}
}}}
\vskip1cm
\centerline{TABLE IV}
\noindent
Estimates for the Bjorken sum rule for deep inelastic polarized
electron-nucleon scattering.
The $\O(\alpha^3_s)$ (first row) results are compared with the exact (known)
results and the $\O(\alpha^4_s)$ (second row) results are compared with those
of \KS.
Their result should be multiplied by $-$1 to compare with ours.
\bigskip
\centerline{\vbox{\halign{
# &\qquad \hfill # &\qquad \hfill # \cr
       SERIES &        ESTIMATE      &       EXACT        \cr
\noalign{\medskip\hrule\medskip}
              &                      &                    \cr
  $-$ BjpSR     &                      &                    \cr
              &                      &                    \cr
 $N_f=3$      &      $-$12.8(6.4)    &       $-$20.2      \cr
              &      $-$112(33)      &       $-$130$\,\,$(\KS)\cr
              &                      &                    \cr
 $N_f=4$      &      $-$10.6(4.4)    &       $-$13.9      \cr
              &      $-$58.2(15.2)   &       $-$68.1$\,\,$(\KS) \cr
              &                      &                    \cr
$N_f=5$       &      $-$8.5(3.4)     &       $-$7.8       \cr
              &      $-$21.1(3.4)    &       $-$17.8$\,\,$(\KS) \cr
\noalign{\medskip\hrule\medskip}
}}}}
\vfill
}
\eject
\bigskip

\vbox{
\centerline{TABLE V}
\noindent
Estimates for $R(\MS)$ for various $N_f$ at the four- (first row)
and five-loop (second row) order.
The four-loop results are compared with the exact (known) results
and the five-loop results are compared with those of \KS.
\bigskip
\centerline{\vbox{\halign{
# &\qquad \hfill # &\qquad \hfill # \cr
SERIES     &         ESTIMATE         &   EXACT             \cr
\noalign{\medskip\hrule\medskip}
           &                          &                     \cr
R RATIO    &                          &                     \cr
           &                          &                     \cr
$N_f = 3$  &         $-14.1\pm2.0$    &      $-$10.27       \cr
           &         $-119.3\pm6.3$   &      $-$128.4$\,\,$(\KS)\cr
           &                          &                     \cr
$N_f = 4$  &         $-12.2\pm1.4$    &      $-$11.5        \cr
           &         $-115.2\pm5.4$   &      $-$111.8$\,\,$(\KS)\cr
           &                          &                     \cr
$N_f = 5$  &         $-10.07\pm1.99$  &      $-$12.76       \cr
           &         $-86.5\pm10.8$   &      $-$96.8$\,\,$(\KS) \cr
\noalign{\medskip\hrule\medskip}
}}} }
\endpage
\bye
\centerline{\psfig{figure=dfuncfig.ps,height=5in}}
\vskip1cm
\vbox{\tenpoint\baselineskip=14pt
\noindent
Relative errors in the $[N/M]$ Pad\'e approximants:
(a) to the
QCD vacuum polarization D-function, evaluated to all orders
in the large-$N_f$ approximation\refmark{\refIXa} --
the rate of convergence agrees with expectations for
a series with a discrete set of Borel poles,
and
(b) to the Borel transform of the D-function series,
where the convergence is particularly striking.
}

\end